\documentclass[12pt,aps,nofootinbib,showpacs,preprint,preprintnumbers]{revtex4}
\usepackage{epsfig}

\def\barD{\overline D{}^0}

\def\barB{\overline B{}^0}
\def\Hbar{\overline{H}}

\def\beq{\begin{equation}}
\def\eeq{\end{equation}}
\def\bea{\begin{eqnarray}}
\def\eea{\end{eqnarray}}

\newcommand{\bra}[1]{\langle #1|}
\newcommand{\ket}[1]{|#1\rangle}

\newcommand{\bld}[1]{\mbox{\boldmath$#1$}}

\begin{document}

\preprint{\vbox {\hbox{WSU-HEP-0502}
\hbox{\tt hep-ph/0506141}}}

\vspace*{1cm}

\title{\bld{$X(3872)$}: Hadronic Molecules in Effective Field Theory}

\author{Mohammad T.~AlFiky$^1$, Fabrizio Gabbiani$^1$, and Alexey A.~Petrov$^{1,2}$}

\affiliation{$^1$Department of Physics and Astronomy, Wayne State University,
Detroit, Michigan 48201\\
$^2$Michigan Center for Theoretical Physics, University of Michigan,
Ann Arbor, Michigan 48109}

\vspace*{2truecm}

\begin{abstract}
We consider the implications from the possibility that the recently observed state
$X(3872)$ is a meson-antimeson molecule. We write an effective Lagrangian
consistent with the heavy-quark and chiral symmetries needed to describe $X(3872)$.
We claim that if $X(3872)$ is a molecular bound state of $D^{*0}$ and $\barD$
mesons, the heavy-quark symmetry requires the existence of the molecular
bound state $X_b$ of $B^{*0}$ and $\barB$ with the mass of 10604 MeV.
\end{abstract}


\maketitle

\section{Introduction}

The past two years have seen several experimental observations of new
heavy and light quark states. Most of
those states contain a charm quark, so their observation rekindled
interest in heavy-flavor spectroscopy~\cite{Petrov:2003un}. The unusual
properties of those states invited some speculations regarding their possible
non-$q\bar q$ nature. Among those is the
$X(3872)$ state which, being discovered in the decay
$X(3872) \to J/\psi \pi^+\pi^-$, contains charm-anticharm
quarks~\cite{MolExp,Reviews}.
While a traditional $c\overline{c}$ quarkonium interpretation of this
state has been proposed~\cite{Barnes:2003vb}, its somewhat unusual mass and
decay patterns prompted a series of more exotic
interpretations~\cite{IntX}. Since the mass of the $X(3872)$ state lies
tantalizingly close to the $D^{*0}\barD$ threshold of 3871.3~MeV, it is
tempting to assume that $X(3872)$ could be a $D^{*0}\barD$ molecular
state~\cite{MoleculeX,Braaten}. Recent Belle data appear to be consistent with
this assignment, preliminarily confirming its $J^{PC}$ = $1^{++}$
quantum numbers~\cite{Abe:2005ix}.
Of course, states of different ``nature''
can mix, if they have the same quantum numbers, further complicating
the interpretation of the experimental data~\cite{Browder:2003fk}.

An unambiguous identification of this state must be done with
many different measurements of its decay and production
patterns. Regardless of whether $X(3872)$ is identified to be a
molecule or a regular $q\bar q$ charmonium, a theoretical analysis of
heavy-meson molecular states should be performed. Until
recently~\cite{Braaten} these studies were done mostly with the
help of various quark models~\cite{MoleculeX,OldStudies}.
In this paper we shall study those states using the techniques of
effective field theories.

This study is possible due to the multitude of scales present in QCD.
The extreme smallness of the binding energy
\beq\label{bindex}
E_b=(m^{\phantom{l}}_{D^0}+m^{\phantom{l}}_{D^{0*}})-M^{\phantom{l}}_X=
-0.6 \pm 1.1~\mbox{MeV}
\eeq
suggests that this state can play the role of the ``deuteron''
(or ``deuson,'' see N.~A.~T\"orn\-qvist's paper in~\cite{MoleculeX})
in meson-meson interactions. The ``deuteron-like'' nature of this
state allows us to use methods similar to those developed for the
description of the deuteron, with the added benefit of heavy-quark
symmetry. The tiny binding energy of this molecular state introduces
an energy scale which is much smaller than the mass of the lightest
particle, the pion, whose exchange can provide binding.
Thus, for a suitable description of this state in the framework of
effective field theory, the pion, along with other particles providing
possible binding contributions (i.e. the $\rho$-meson and other higher mass
resonances), must be integrated out. The resulting Lagrangian should
contain only heavy-meson degrees of freedom with interactions
approximated by local four-boson terms constrained only by the
symmetries of the theory. This approach is similar to
the description of the deuteron in the effective theory without
dynamical pions~\cite{Weinberg}. Nevertheless, we shall often appeal
to the ``exchange picture'' to gain insight into the structure of the
effective Lagrangian.

This approach provides a model-independent description of molecular
states with somewhat limited predictive power. In particular, we would
not be able to say {\it whether} the state $X(3872)$ is a $D^{*0}\barD$
molecule or not. What we {\it would} be able to say is that if indeed
$X(3872)$ is a $D^{*0}\barD$ molecule, the heavy-quark symmetry makes
a definite statement on the existence of a molecular state in the
$B^{*0}\barB$ system. We also show that even though $D$ and $D^*$
are degenerate in the heavy-quark limit, the existence of a
molecular state in $D^{*0}\barD$ channel does not necessarily
imply a molecular state in the $D^0\overline {D}^0$ or $B^{*0}\barD$
channels.

This paper is organized as follows. In Sec. II we write the most
general effective Lagrangian consistent with the heavy-quark and chiral
symmetry. In Sec. III we obtain the bound-state energy by solving
a system of Lippmann-Schwinger equations and relate the bound-state
energies of $D^{*0}\barD$ and $B^{*0}\barB$ states.
We present our conclusions in Sec. IV.

\section{The Effective Lagrangian}

In order to describe the molecular states of heavy mesons we need an effective Lagrangian
which contains two- and four-body interaction terms.
The two-body effective Lagrangian that describes the strong interactions of
the heavy mesons $P$ and $P^*$ ($P=B,D$) containing one heavy quark $Q$
is well known~\cite{Grinstein:1992qt}:
\bea\label{Lagr2}
{\cal L}_2 ~&=& ~-i \mbox{Tr} \left[ \Hbar^{(Q)} v \cdot D H^{(Q)} \right]
- \frac{1}{2 m^{\phantom{l}}_P} \mbox{Tr} \left[ \Hbar^{(Q)} D^2 H^{(Q)} \right]
\nonumber \\
&+& ~\frac{\lambda_2}{m^{\phantom{l}}_P}  \mbox{Tr}
\left[ \Hbar^{(Q)} \sigma^{\mu\nu } H^{(Q)} \sigma_{\mu\nu} \right]
+ \frac{ig}{2} \mbox{Tr}  \Hbar^{(Q)} H^{(Q)} \gamma_\mu \gamma_5
\left[\xi^\dagger \partial^\mu \xi - \xi \partial^\mu \xi^\dagger
\right] + ...
\eea
where the ellipsis denotes terms with more derivatives or including explicit
factors of light quark masses, $D_{ab}^\mu=
\delta_{ab}\partial^\mu-(1/2)\left(\xi^\dagger \partial^\mu \xi +
\xi \partial^\mu \xi^\dagger
\right)_{ab}$, and $g$ is the $P^{*}P\pi$ coupling.
As usual, we introduce a superfield describing the combined doublet of
pseudoscalar heavy-meson fields $P_a = \left(P^0, P^+\right)$ and
their vector counterparts with $v\cdot P^{*(Q)}_{a}=0$,
\beq
H_a^{(Q)}=\frac{1+\not{v}}{2}\left[
P^{*(Q)}_{a\mu} \gamma^\mu - P_a^{(Q)} \gamma_5
\right], \qquad \overline{H}^{(Q) a} = \gamma^0 H_a^{(Q)\dagger} \gamma^0.
\eeq
These fields have the usual transformation properties under the heavy-quark
spin symmetry and SU(2)$_V$ flavor symmetry\footnote{A generalization of this
discussion to the flavor SU(3) symmetry is rather straightforward.},
\beq
H_a^{(Q)} \to S\left(H^{(Q)}U^\dagger\right)_a, \qquad \overline{H}^{(Q) a} \to
\left(U \overline{H}^{(Q)}\right)^a S^{-1},
\eeq
and describe heavy mesons with a definite velocity $v$~\cite{Falk:1991nq}.
The third term in Eq.~(\ref{Lagr2}) is needed to account for the $P-P^*$ mass
difference $\Delta\equiv m^{\phantom{l}}_{P^*}-m^{\phantom{l}}_P=-2\lambda_2/m^{\phantom{l}}_P$.

The pseudo-Goldstone fields are introduced as $\xi=e^{i\widetilde{M}/f}$,
where $\widetilde{M}$ is the usual meson matrix~\cite{Donoghue:1992dd}
\beq
\widetilde{M}=\left(
\begin{array}{cc}
\frac{1}{\sqrt{2}}\pi^0 & \pi^+ \\
\pi^- & -\frac{1}{\sqrt{2}}\pi^0
\end{array}
\right),
\eeq
and $f\simeq135$ MeV is the pion decay constant. Notice that since
heavy quark-antiquark pair production is absent in this effective theory,
the effective Lagrangian of Eq.~(\ref{Lagr2}) does not contain heavy
antimeson degrees of freedom. Since we are describing the molecular
states of heavy mesons, those fields should have to be explicitly added to
the Lagrangian. The fields $H_a^{(\overline{Q})}$ and $H_a^{(\overline{Q})\dagger}$
that describe the propagation of heavy antimesons, i.e.
containing the heavy antiquark $\overline{Q}$, are introduced as
\beq
H_a^{(\overline{Q})}=\left[
P^{*(\overline{Q})}_{a\mu} \gamma^\mu - P_a^{(\overline{Q})} \gamma_5
\right] \frac{1-\not{v}}{2}, \qquad
\overline{H}^{(\overline{Q}) a} = \gamma^0 H_a^{(\overline{Q})\dagger} \gamma^0,
\eeq
and transform as $H_a^{(\overline{Q})} \to
\left(U H^{(\overline{Q})}\right)_a S^{-1}$ and
$\overline{H}^{(\overline{Q}) a} \to
S \left(\overline{H}^{(\overline{Q})}U^{\dagger}\right)^a$ under heavy-quark spin
and SU(2)$_V$ symmetry.

In order to write an effective Lagrangian describing the properties of
$X(3872)$, we need to couple the fields $H_a^{(Q)}$ and $H^{(\overline{Q})a}$ so that
the resulting Lagrangian respects the heavy-quark spin and chiral symmetries.
Since the binding energy of $X(3872)$ is small, the size of a bound state is rather
large. This means that the particular details of the interaction of
the heavy meson and antimeson pair (for example, a $\rho$-meson exchange
contribution) are irrelevant for the description of such a bound state and can be well
approximated by four-meson local interactions. One can write a Lagrangian describing
$X(3872)$ by first writing an effective Lagrangian above $\mu=m_\pi$ and then matching it
onto the Lagrangian for $\mu<m_\pi$, i.e. integrating out the pion degrees of freedom.

The general effective Lagrangian consistent with heavy-quark spin and chiral symmetries
can be written as
\beq\label{Lagr}
{\cal L}={\cal L}_2+{\cal L}_4,
\eeq
where the two-body piece, consistent with reparametrization invariance, is given
by Eq.~(\ref{Lagr2}) and the four-body piece is
\bea\label{Lagr4}
-{\cal L}_4&=& \frac{C_1}{4} \mbox{Tr} \left[ \Hbar^{(Q)} H^{(Q)} \gamma_\mu \right]
\mbox{Tr} \left[ H^{(\overline{Q})} \Hbar^{(\overline{Q})} \gamma^\mu \right]
+ \frac{C_2}{4} \mbox{Tr} \left[ \Hbar^{(Q)}  H^{(Q)} \gamma_\mu \gamma_5 \right]
\mbox{Tr} \left[ H^{(\overline{Q})} \Hbar^{(\overline{Q})} \gamma^\mu \gamma_5 \right].
\eea
This Lagrangian, together with ${\cal L}_2$, describes the scattering of $P$ and $P^*$ mesons
at the energy scale above $m_\pi$. Integrating out the pion degrees of freedom at tree
level corresponds to a modification $C_2' \to C_2 + (2/3) \left(g/f\right)^2$. Since in this paper
we will not discuss matching at higher orders, the Lagrangian in Eq.~(\ref{Lagr4}) will be
used for the calculation of the bound state properties of $X(3872)$.

By virtue of the heavy-quark spin symmetry, the same Lagrangian governs the four-boson
interactions of {\it all} $P_a^{(*)}=D^{(*)}$ or $B^{(*)}$
states, while the flavor $SU(2)_V$ implies that there could be four such states
for each $P_a^{(*)}\overline{P}_b^{(*)}$. Indeed, not
all of these states are bound. Here we shall concentrate on $X(3872)$, which is a bound
state of two {\it neutral} bosons, $P_a\equiv P^0\equiv P$, assuming
the isospin breaking advocated
in Ref.~\cite{MoleculeX}. Notice that the most general Lagrangian involves two couplings,
$C_1$ and $C_2$. Other Dirac structures are possible, but will yield the same
Lagrangian for the $P\overline{P^*}$ bound state.

In order to relate the properties of $P\overline{P^*}$ molecules in the charm and beauty sectors
we shall need to see how $C_i$ couplings scale as functions of the heavy-quark mass $M$.
To see this, we recall that a system of two heavy particles requires
a nonrelativistic $v/c$ expansion, not a $1/M$ expansion. This is
necessary to avoid that the resulting loop integrals acquire pinch
singularities~\cite{Weinberg}.
Therefore, we must powercount $p^0 \sim \vec{p}^2/M$, where $\vec{p}$ is a characteristic 
3-momentum of a heavy meson in the $P\overline{P^*}$ molecular bound state, which implies 
that the first and the second
terms in Eq.~(\ref{Lagr2}) scale in the same way. Since the action $S=\int d^4x ~{\cal L}$ does
not scale with the heavy-quark mass, this implies that $d^4x \sim M$ and the
Lagrangian density ${\cal L} \sim 1/M$. The kinetic term in Eq.~(\ref{Lagr2}) then gives the
expected scaling of the heavy-meson field $H\sim P \sim P^*\sim M^0$, which in turn
implies from Eq.~(\ref{Lagr4}) that the couplings
\beq\label{Scaling}
C_i \sim 1/M.
\eeq
This dimensional analysis, however, cannot be used to predict the relative contributions of 
other couplings in ${\cal L}_4$, say, relativistic corrections to Eq.~(\ref{Lagr4}), because
of the fine-tuning which produces a molecular state close to threshold in the first 
place~\cite{Braaten,Weinberg}. We will use it only to relate properties of $DD^*$ 
and $BB^*$ systems. Similar dimensional analysis was proposed for non-relativistic 
QCD~\cite{Luke:1996hj}.

Evaluating the traces yields for the $P\overline{P^*}$ sector
\bea\label{LocalLagr}
{\cal L}_{4,PP^*} = &-& C_1 P^{(Q)\dagger} P^{(Q)}
P^{*(\overline{Q})\dagger}_\mu P^{* (\overline{Q}) \mu}
- C_1 P^{*(Q)\dagger}_\mu P^{*(Q) \mu}
P^{(\overline{Q})\dagger} P^{(\overline{Q})} \nonumber \\
&+& C_2 P^{(Q)\dagger} P^{*(Q)}_\mu
P^{* (\overline{Q})\dagger \mu} P^{(\overline{Q})}
+ C_2 P^{* (Q)\dagger}_\mu P^{(Q)}
P^{(\overline{Q})\dagger} P^{* (\overline{Q}) \mu}
+\dots
\eea
Notice that this Lagrangian differs from the one used in~\cite{Braaten}, where the
interaction strength is described by a single parameter $\lambda=-C_1=C_2$. The difference
can be understood in the ``exchange'' model, where $C_1$ and $C_2$ come from the
exchanges of mesons of different masses and parity. In this language the model of Ref.~\cite{Braaten}
corresponds to the situation of degenerate parity states. In QCD, however,
negative parity states are generally lighter than their positive parity counterparts.
This is especially clear for the lightest octet of pseudoscalar mesons, where chiral symmetry
forces their masses to be almost zero, while all the corresponding scalar mesons have
masses of the order of 1 GeV. We will nevertheless show that the resulting binding energy 
still depends on a single parameter, a linear combination of $C_1$ and $C_2$.

Similarly, one obtains the component Lagrangian governing the interactions of
$P$ and $\overline P$,
\beq\label{LocalLagrPP}
{\cal L}_{4,PP} = C_1 P^{(Q)\dagger} P^{(Q)}
P^{(\overline{Q})\dagger} P^{(\overline{Q})}.
\eeq
Clearly, one cannot relate the existence of the bound state in the
$P\overline{P^*}$ and $P\overline{P}$ channels, as the properties of
the latter will depend on $C_1$ alone, not a linear combination of $C_1$ and $C_2$.

\section{Properties of bound states}

In order to describe bound states we shall modify the approach of
S.~Weinberg~\cite{Weinberg}. The lowest-energy bound state of $P$ and
$\overline{P^*}$ is an eigenstate of charge conjugation. Here we carry out
our calculation for ($P$,$\overline{P^*}$) = ($D$,$\overline{D^*}$),
but analogous considerations will apply to the $B$ system. The two
eigenstates of charge conjugation will be given by
\beq\label{Eigenstate}
\ket{X_{\pm}}=\frac{1}{\sqrt{2}}\left[
\ket{D^* \overline{D}} \pm \ket{D \overline{D}^*}
\right].
\eeq
To find the bound-state energy of $X(3872)$ with $J^{PC}=1^{++}$,
we shall look for a pole of the transition amplitude $T_{++}=\bra{X_+}T\ket{X_+}$.

\begin{figure}[tb]
\centerline{\epsfxsize=12cm\epsffile{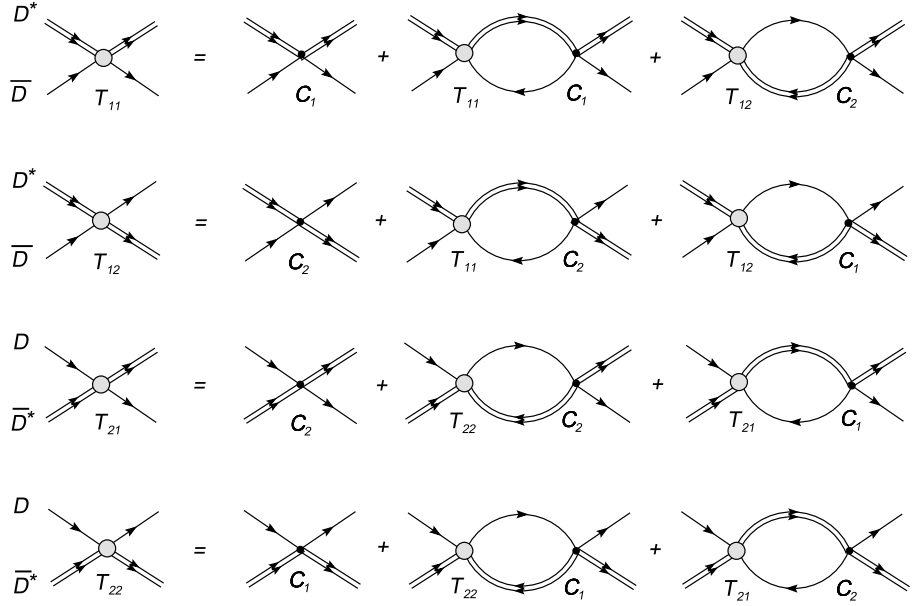}}
\centerline{\parbox{17cm}{\caption{\label{fig1}
Transition amplitudes for the $D-\overline{D}^*$ scattering written in the form
of Lippmann-Schwinger equations. Double lines indicate the vector $D^*$ or
$\overline{D}^*$ mesons, solid lines pseudoscalar $D$ or $\overline{D}$ states.}}}
\end{figure}
We first define the following transition amplitudes,
\bea\label{Ts}
T_{11}&=&\langle D^* \overline{D}| T | D^* \overline{D} \rangle, \quad
T_{12}=\langle D^* \overline{D}| T | D \overline{D}^* \rangle,
\nonumber \\
T_{21}&=&\langle D \overline{D}^*| T | D^* \overline{D} \rangle, \quad
T_{22}=\langle D \overline{D}^*| T | D \overline{D}^* \rangle,
\eea
which correspond to the scattering of $D$ and $D^*$ mesons. Clearly,
at tree level, $T_{ii} \sim C_1$ and $T_{ij,~i\neq j} \sim C_2$, since
we consider only contact interactions. But we also have to include a
resummation of loop contributions to complete the leading
order~\cite{Weinberg}. These transition amplitudes satisfy a system of
Lippmann-Schwinger equations depicted in Fig.~\ref{fig1},
\bea\label{LSE}
i T_{11}&=& -i C_1 + \int \frac{d^4 q}{(2\pi)^4} T_{11}
G_{PP^*} C_1 - \int \frac{d^4 q}{(2\pi)^4} T_{12}
G_{PP^*} C_2,
\nonumber \\
i T_{12}&=& ~~i C_2 - \int \frac{d^4 q}{(2\pi)^4} T_{11}
G_{PP^*} C_2 + \int \frac{d^4 q}{(2\pi)^4} T_{12}
G_{PP^*} C_1,
\nonumber \\
i T_{21}&=& ~~i C_2 + \int \frac{d^4 q}{(2\pi)^4} T_{21}
G_{PP^*} C_1 - \int \frac{d^4 q}{(2\pi)^4} T_{22}
G_{PP^*} C_2,
\\
i T_{22}&=& -i C_1 - \int \frac{d^4 q}{(2\pi)^4} T_{21}
G_{PP^*} C_2 + \int \frac{d^4 q}{(2\pi)^4} T_{22}
G_{PP^*} C_1,
\nonumber
\eea
where
\beq\label{Tra}
G_{PP^*}=
 \frac{1}{4} \frac{1}{\left(\vec{p}^2/{2m^{\phantom{l}}_{D^*}}+
q_0-\Delta-\vec{q}^2/{2m^{\phantom{l}}_{D^*}}+i\epsilon\right)
\left(\vec{p}^2/{2m^{\phantom{l}}_{D}}-q_0-\vec{q}^2/{2m^{\phantom{l}}_D}+i\epsilon \right)},
\eeq
$\vec{p}$ is the momentum of one of the mesons in the
center-of-mass system, and we canceled out factors of
$m^{\phantom{l}}_D m^{\phantom{l}}_{D^*}$ appearing on both sides of Eq.~(\ref{LSE}).
Note that in Eq.~(\ref{Tra}) the vector propagator includes the mass
difference $\Delta$, as a consequence of the term proportional to
$\lambda_2$ in the Lagrangian of Eq.~(\ref{Lagr2})~\cite{Grinstein:1992qt}.
This choice of propagators is not unique, but our results will not
depend on it, because it amounts to a choice of a finite phase
multiplying the heavy-meson fields. This rephasing is equivalent to
measuring energies with respect to the pseudoscalar mass, $m_D$.
Then, the position of the transition amplitude pole
on the energy scale will be measured with respect to the ``constituent mass'' of the system,
which is in our case twice the pseudoscalar mass $m^{\phantom{l}}_D$~\cite{Manohar:2000dt}.
A different choice of phase will give different propagators
but also a different ``constituent mass''.

Since we are interested in the pole of the amplitude $T_{++}$, we must
diagonalize this system of equations rewritten in an algebraic matrix form,
\bea\label{LSEMatrix}
\left(
\begin{array}{c}
T_{11} \\
T_{12} \\
T_{21} \\
T_{22}
\end{array}
\right)
=
\left(
\begin{array}{c}
-C_1 \\
C_2 \\
C_2 \\
-C_1
\end{array}
\right)+
i\widetilde{A} \left(
\begin{array}{cccc}
-C_1 & C_2 & 0 & 0 \\
C_2 & -C_1 & 0 & 0 \\
0 & 0 & -C_1 & C_2 \\
0 & 0 & C_2 & -C_1
\end{array}
\right)
\left(
\begin{array}{c}
T_{11} \\
T_{12} \\
T_{21} \\
T_{22}
\end{array}
\right).
\eea
Notice that the matrix is in the block-diagonal form, which allows us to solve
Eq.~(\ref{LSEMatrix}) in two steps working only with $2\times 2$ matrices. The solution
of Eq.~(\ref{LSEMatrix}) produces the $T_{++}$ amplitude,
\beq\label{Solution}
T_{++}=\frac{1}{2}\left( T_{11}+T_{12}+T_{21}+T_{22} \right)=
\frac{\lambda}{1-i\lambda \widetilde{A}},
\eeq
with $\lambda=-C_1+C_2$, and $\widetilde{A}$ is a (divergent) integral
\bea\label{Integral}
\widetilde{A}&\;=\;&\frac{1}{4}\int\frac{d^4q}{(2\pi)^4}
\frac{1}{\left(\vec{p}^2/{2m^{\phantom{l}}_{D^*}}+
q_0-\Delta-\vec{q}^2/{2m^{\phantom{l}}_{D^*}}+i\epsilon\right)
\left(\vec{p}^2/{2m^{\phantom{l}}_D}-q_0-\vec{q}^2/{2m^{\phantom{l}}_D}+i\epsilon \right)}
\nonumber \\
&\;=\;& \frac{i}{4} 2 \mu^{\phantom{l}}_{DD^*}
\int \frac{d^3q}{(2\pi)^3} \frac{1}{\vec{q}^2-2\mu^{\phantom{l}}_{DD^*}\left(E-\Delta\right)
-i\epsilon},
\eea
where $E=\vec{p}^2/2\mu^{\phantom{l}}_{DD^*}$, $\mu^{\phantom{l}}_{DD^*}$ is the reduced mass of
the $DD^*$ system, and we have used the residue theorem to evaluate the integral over $q^0$.
The divergence of the integral of Eq.~(\ref{Integral}), as usual, is removed by
renormalization. We choose to define a renormalized $\lambda^{\phantom{l}}_R$ within the $MS$
subtraction
scheme in dimensional regularization. In this scheme the integral $\widetilde{A}$ is
finite, which corresponds to an implicit subtraction of power divergences in
Eq.~(\ref{Integral}). Computing the second integral in Eq.~(\ref{Integral})
by analytically continuing to $d-1$ dimensions yields
\bea
\widetilde{A}= -\frac{1}{8 \pi} \mu^{\phantom{l}}_{DD^*}
|\vec{p}| \sqrt{1-\frac{2 \mu^{\phantom{l}}_{DD^*}\Delta}{\vec{p}^2}}.
\eea
This implies for the transition amplitude
\bea\label{FinAmp}
T_{++}=
\frac{\lambda^{\phantom{l}}_R}{1+(i/{8\pi})\lambda^{\phantom{l}}_R\, \mu^{\phantom{l}}_{DD^*}
|\vec{p}|
\sqrt{1-2 \mu^{\phantom{l}}_{DD^*}\Delta/{\vec{p}^2}}}.
\eea
The position of the pole of the molecular state on the energy scale
can be read off Eq.~(\ref{FinAmp}),
\beq\label{Pole}
E_{\rm Pole}=\frac{32 \pi^2}{\lambda_R^2 \mu_{DD^*}^3}-\Delta.
\eeq
This is the amount of energy we must subtract from the ``constituent mass''
of the system, determined above as $2 m^{\phantom{l}}_D$,
in order to calculate the mass
\beq
M^{\phantom{l}}_X=2 m^{\phantom{l}}_D-E_{\rm Pole}=2
m^{\phantom{l}}_D+\Delta-\frac{32 \pi^2}{\lambda_R^2 \mu_{DD^*}^3}.
\eeq
Recalling the definition of binding energy, Eq.~(\ref{bindex}),
and that $m^{\phantom{l}}_{D^*}$ = $m^{\phantom{l}}_{D}$ +  $\Delta$,
we infer
\beq\label{Binding}
E_b=\frac{32 \pi^2}{\lambda_R^2 \mu_{DD^*}^3}.
\eeq
Assuming $E_b$ = 0.5 MeV, which is one sigma below the central value in
Eq.~(\ref{bindex}) \cite{MolExp}, and the experimental values for the
masses~\cite{PDG}, we obtain
\beq\label{Lambda}
\lambda^{\phantom{l}}_R \simeq 8.4 \times 10^{-4} \ {\rm MeV}^{-2}.
\eeq
Note that the binding energy scales as $1/M$ in the
heavy-quark limit. Thus, the smallness of the binding energy is implied in the
heavy-quark limit. The small binding energy of the $X(3872)$
state implies that the scattering length
$a^{\phantom{l}}_D$ is large and can be written as
\beq
a^{\phantom{l}}_D=\sqrt{\left(2 \mu^{\phantom{l}}_{DD^*} E_b \right)^{-1}}=
\frac{\lambda^{\phantom{l}}_R \mu^{\phantom{l}}_{DD^*}}
{8 \pi},
\eeq
yielding a numerical value $a^{\phantom{l}}_D$ = 6.3 fm.
Since the scattering length is large, universality implies that the
leading-order wave function of $X(3872)$ is known,
\beq
\psi^{\phantom{l}}_{DD^*}(r) =\frac{e^{-r/a^{\phantom{l}}_D}}{\sqrt{2\pi a^{\phantom{\dagger}}_D}r}.
\eeq
This can be used to predict the production and decay properties of
$X(3872)$~\cite{Braaten,ExpX}.

Once we establish the molecular nature of $X(3872)$, its experimental
mass gives us its binding energy. The latter is dependent on the coupling
constant $\lambda^{\phantom{l}}_R$, and may be used to predict the binding energies of
hypothetical molecular bound states in the beauty sector,
as well as to discuss its implications for the beauty-charm sector. Alternatively, the coupling
constant $\lambda^{\phantom{l}}_R$ can be fixed from the resonance-exchange model,
in which case (model-dependent) predictions are possible for all the heavy sectors.
Taking into account the scaling of $\lambda^{\phantom{l}}_R$ with $M$ given by
Eq.~(\ref{Scaling}), we obtain
\beq\label{Rescale}
\lambda_R^B \sim \lambda_R^D \frac{\mu^{\phantom{l}}_{DD^*}}{\mu^{\phantom{l}}_{BB^*}}.
\eeq
Formula (\ref{Binding}) can now be used for the $B$ system.
This implies the existence of an S-wave bound states with
binding energy $E_b$ = 0.18 MeV, mass $M_{X_b}$ = 10604 MeV, and a
scattering length $a^{\phantom{l}}_B$ = $a^{\phantom{l}}_D$ = 6.3 fm,
because of the lack of scaling of the scattering length with the
heavy-quark mass. The above prediction for the
binding energy is lower than the quark-model prediction of Wong
in~\cite{MoleculeX}. Similar considerations apply to $D^0 \barD$ and $B^0 \barB$ states:
In their case the starting point is the
Lagrangian term in Eq.~(\ref{LocalLagrPP}). Since it involves only a
single term, the calculations are actually easier and involve only
one Lippmann-Schwinger equation. The resulting binding energy is then
\beq\label{BindingC}
E_b=\frac{256 \pi^2}{C_{1R}^2 m_D^3},
\eeq
and the equivalent formula for the $B^0 \barB$ system may be obtained
by rescaling the coupling constant $C_{1R}^2$ as we did with
$\lambda^{\phantom{l}}_R$ in Eq.~(\ref{Rescale}).
Examining Eq.~(\ref{BindingC}) we immediately notice that the existence of
a bound state in the $D^*\overline{D}$ channel does not dictate the properties
of a possible bound state in the $D^0 \barD$ or $B^0 \barB$ channels, since $C_1$ and
$C_2$ are generally not related to each other.

The discussion of the beauty-charm system parallels what was done above. In this case the situation is
a bit different, because there are two states, $B^0 \overline{D}^{*0}$ and $D^0 \overline{B}^{*0}$.
The treatment of these states depends on how the heavy-quark limit is taken.
They have different masses, since $\Delta_{BB^*}\neq\Delta_{DD^*}$.
This implies that these two states do not mix and have to be treated
separately, so that their binding energies could be obtained by respectively
substituting $\lambda^{\phantom{l}}_R$ $\to$ $C_{1R}$ and $\mu^{\phantom{l}}_{DD^*}$ $\to$
$\mu^{\phantom{l}}_{BD^*}$ or $\mu^{\phantom{l}}_{DB^*}$ in Eq.~(\ref{Binding}). But just like in
the case of $D^0 \barD$ or $B^0 \barB$ channels, we cannot
predict bound states in these channels from the heavy-quark
symmetry arguments alone.

\section{Conclusion}

We have used an effective field theory approach in the analysis of the
likely molecular state $X(3872)$, by describing its binding
interaction with contact terms in a heavy-quark symmetric
Lagrangian. The flexibility of this description allows us to ignore the
details of the interaction and to concentrate on its effects, namely a
shallow bound state and a large scattering length. Taking into account
the universality and the scaling of the effective coupling constants
we are able to extend our description to the $B$ system.
We found that if $X(3872)$ is indeed a molecular bound state of $D^{*0}$ and
$\barD$ mesons, the heavy-quark symmetry requires the existence of the molecular
bound state $X_b$ of $B^{*0}$ and $\barB$ with the mass of 10604 MeV.

\section*{ACKNOWLEDGMENTS}

A.~P. thanks T.~Mehen and R.~Hill for useful conversations.
F.~G. thanks P.~Bedaque for his clarifying remarks.
The authors would like to thank E.~Braaten for reading the manuscript and for
helpful comments. A.~P. also thanks the Institute for Nuclear Theory at the
University of Washington for its hospitality and the Department of Energy for
partial support during the completion of this work.
This work was supported in part by the U.S.\ National Science Foundation under
Grant PHY--0244853, and by the U.S.\ Department of Energy under Contract
DE-FG02-96ER41005.



\end{document}